\documentclass[prd, superscriptaddress,nofootinbib,showpacs]{revtex4}
\usepackage{graphicx}
\usepackage{dcolumn}
\usepackage{slashbox,multirow}
\usepackage{tabularx}
\usepackage{amsmath,amsthm,amssymb}
\usepackage{cases}
\usepackage{epsfig}
\begin{document}

\title{Late-time tails of self-gravitating skyrmions}

\author{Stanis{\l}aw Zaj\c{a}c}
\affiliation{H. Niewodniczanski Institute of Nuclear
   Physics, Polish Academy of Sciences,  Krak\'ow, Poland}
\date{\today}
\begin{abstract}
We consider the long-time behaviour of spherically symmetric solutions in the Einstein-Skyrme model.
Using $nonlinear$ perturbation analysis we obtain the leading order estimation of the tail in the topologically
trivial sector ($B=0$) of the model. We show that solutions starting from small compactly supported
initial data decay as $t^{-4}$ at future timelike infinity and as $u^{-2}$ at future null infinity. 
We also verified that long-time behaviour for the tail in Einstein-Skyrme model is exactly the same as it was obtained for wave maps.

\end{abstract}

\maketitle

\section{Introduction}
This paper concerns the late-time asymptotic behaviour of a spherically symmetric self-gravitating
Einstein-Skyrme (ES) model. It is an extension of the paper \cite{SZ} where we studied quasinormal modes in intermediate asymptotics. It is also an extension of work done in \cite{BCR1} where the expression for the tail in flat space was obtained. The results of this paper are closely connected to the results of paper \cite{BCRZ} where the evolution of wave maps was studied. As we remarked in paper \cite{SZ}, in gravitating Skyrme model the linear perturbation method predicts power-law index $\gamma=5$ for the tail. This estimation is in clear conflict with early numerical results for the tails in ES \cite{TCh} which were later confirmed by the results of paper \cite{SZ}. To explain this disagreement we have studied the expression for the tail in a gravitating wave maps model for details see paper \cite{BCRZ}  where we expected similar long-time asymptotics as for the Skyrme model. In the current paper direct calculations in gravitating Skyrme model are performed.

In the self--gravitating Skyrme model the most interesting problem is certainly the description of the relaxation to the static Skyrmion. Unfortunately, due to the lack of analytic formulae describing static Skyrme soliton, the description of this problem is very tedious. To avoid these difficulties  we follow in the same way as was done by Bizo\'n et al. \cite{BCR1} so that we study the relaxation to the vacuum in the topologically trivial $B=0$ sector. To estimate the parameters of the tail we apply perturbation techniques elaborated in \cite{{BCR2},{BCR3},{BCR4},{BCR5}}. Using these techniques we will demonstrate  that the third-order expression for the tail agrees perfectly with numerical results for small initial data. The plan of this paper is as follows. In section II we remind the  reader the field equations of the model and shortly demonstrate the iterative scheme. We present the difference between Einstein-Skyrme model and wave maps model which was analysed in paper \cite{BCRZ} for $\ell=1$. In the last section we demonstrate the numerical evidence confirming our analytical estimations for the tails.

%

\section{Theoretical background}
We consider the Einstein--Skyrme model with dynamics given by the Lagrangian \cite{Skyrme}:
\begin{equation}\label{matter}
L=\frac{f^{2}}{4}Tr(\nabla_{a}U\nabla^{a}U^{-1})+\frac{1}{32e^{2}}Tr[(\nabla_{a}U)U^{-1},(\nabla_{b}U)U^{-1}]^{2} - \frac{1}{16 \pi G} R\,,
\end{equation}
where $\nabla_{a}$ is the covariant derivative  with respect to the spacetime metric, $G$ -- is gravitational constans and $R$ -- is a scalar of curvature.
We assume spherical symmetry and parametrize the metric as follows:
\begin{equation}\label{metric}
ds^{2}=-e^{-2\delta(r,t)}N(r,t)dt^{2}+N^{-1}(r,t)dr^{2}+r^{2}d\Omega^{2},
\end{equation}
where $d\Omega^{2}$ is a metric on the unit 2--sphere.
Applying the standard hedgehog ansatz $U=exp(i\overrightarrow{\sigma}\cdot\hat{r} F(r,t)),$ where $\overrightarrow{\sigma}$ is the vector of Pauli matrices and $ \hat{r}$ -- unit radial vector, we obtain the following set of $ES$ equations:
\begin{equation} \label{momentum}
\dot{m}=\alpha e^{-\delta}N^{2}PF',
\end{equation}
\begin{equation}\label{hamilton}
m'=\frac{\alpha}{2}\left(2 \sin^{2}F+\frac{\sin^{4}F}{r^{2}}+uN(\frac{P^{2}}{u^{2}}+F'^{2})\right),
\end{equation}
\begin{equation}\label{delta}
\delta'=-\frac{\alpha u}{r}\left(\frac{P^{2}}{u^{2}}+F'^{2}\right),
\end{equation}
\begin{equation}\label{wave}
\dot{P}=(e^{-\delta}NuF')'+ \sin(2F) e^{-\delta}\left(N(\frac{P^{2}}{u^{2}}-F'^{2})-\frac{\sin^{2}F}{r^{2}}-1\right).
\end{equation}
Here $P$ and $u$ are auxiliary variables defined as: $P=ue^{\delta}N^{-1}\dot{F}$ and $u=r^{2}+2\sin^{2}F$, $m(t,r)$ is the mass function defined as: $m(t,r)=\frac{r(1-N)}{2}$ and $\alpha=4\pi Gf^{2}$ is dimensionless coupling constant. The expression for the tail for $\alpha = 0$ (flat space) was obtained in paper \cite{BCR1}; here we consider gravitating case $\alpha > 0$.

To obtain the estimation of the tail we study the evolution of the system described by (\ref{momentum})-(\ref{wave}) starting with small, smooth and compactly supported initial data
\begin{equation}
\label{id} F(0,r) = \varepsilon f(r), \qquad \dot F(0,r) = \varepsilon g(r)\,.
\end{equation}
Following \cite{{BCR2},{BCR3},{BCR4},{BCR5}} we postulate perturbation expansion
\begin{eqnarray}
m(t,r) &=& m_0(t,r) + \varepsilon m_1(t,r) + \varepsilon^2 m_2(t,r) + \dots,
\\
\delta(t,r) &=& \delta_0(t,r) + \varepsilon \delta_1(t,r) + \varepsilon^2 \delta_2(t,r) + \dots,
\\
F(t,r) &=& F_0(t,r) + \varepsilon F_1(t,r) + \varepsilon^2 F_2(t,r) + \varepsilon^3 F_3(t,r) +
\dots . \label{pertexpansion}
\end{eqnarray}
Collecting the terms with the same power of $\varepsilon$ we obtain a set of equations for the expansion functions which we solve recursively. We are studying the relaxation process to the Minkowski space-time, so $m_0=\delta_0=F_0=0$.

In the first order in $\varepsilon$ the requirement of regularity of the metric function $N$  at the origin  and choice of gauge $\delta(t,r=0)=0$ require that $m_1=\delta_1=0$. In this perturbation order we obtain free $\ell=1$ radial wave equation for the $F_1$ function:
\begin{equation}
\label{Box_f1} \Box F_1 =0\,,\qquad
\Box =\partial_t^2-\partial_r^2-\frac{2}{r}\partial_r+\frac{2}{r^2}\,,
\end{equation}
with initial data $F_1(0,r) = f(r), \dot F_1(0,r) =  g(r)$.
The general  regular solution of equation (\ref{Box_f1}) has the form
\begin{equation}
\label{f1} F_1 (t,r) =\frac{a'(t-r)+a'(t+r)}{r} +\frac{a(t-r)-a(t+r)}{r^2}\,,
\end{equation}
where the generating function $a(r)$ is determined by initial data.

In the second perturbation order we obtain the free $\ell=1$ radial wave equation $\Box F_2 = 0$; however, contrary to the previous $F_1(t,r)$ case, the initial data for $F_2$ are zero so $F_2$ has to vanish. In this order of perturbation expansion, the metric functions satisfy the following equations
\begin{eqnarray}
\label{m2p} m'_2 &=& \frac{\alpha}{2}\, r^2 \left( \dot{F}_1^2 + F_1'^2 +
\frac{2}{r^2} F_1^2\right)\,,
\\
\label{m2dot} \dot{m}_2 &=&  \alpha\, r^2\, \dot{F}_1\, F'_1\,,
\\
\label{delta2prime} \delta'_2 &=& - \alpha\, r\, (\,\dot{F}_1^2 + F_1'^2\,)\,.
\end{eqnarray}
Finally, in the third order in $\varepsilon$ we get following equation for $F_3$
\begin{eqnarray}
\label{Box_f3} \Box F_3^{Skyrme} &{=}& -2 \delta_2 \ddot{F}_1 - \dot{\delta}_2 \dot{F}_1- \delta'_2 F'_1 -
\frac{2}{r}\left(m'_2 F'_1 + \dot{m}_2 \dot{F}_1\right)+\,\frac{m_2}{r}\left(\frac{4}{r^2}F_1\,-\frac{6}{r}F'_1\,-4 F''_1\right)
\nonumber\\
&+&  \frac{4}{3 r^2}F_1^3\,+ \textbf{b}\,,
\end{eqnarray}
where
\begin{equation}
\label{b} \textbf{b} = \frac{2}{r^4}\left(F_1^3 - 2 r F_1^2\,F_1' + r^2\,F_1\,(F_1'^2-\dot{F}_1^2)\right)\,.
\end{equation}\,

We may compare this equation for the third-order perturbation $F_3^{Skyrme}(t,r)$ with corresponding equations obtained for wave maps (equation (21) in paper \cite{BCRZ}) and Skyrmion in flat space (equations (17)-(18) in paper \cite{BCR2}). To make this comparison simpler we rewrite the expression for the wave maps in metric parametrization which we used in this paper (see equation (\ref{metric})). For $\ell=1$ this expression reads:
\begin{eqnarray}
\label{Box_f3wave} \Box F_{3}^{wave-map} &{=}& -2 \delta_2 \ddot{F}_1 - \dot{\delta}_2 \dot{F}_1- \delta'_2 F'_1 -
\frac{2}{r}\left(m'_2 F'_1 + \dot{m}_2 \dot{F}_1\right)+\,\frac{m_2}{r}\left(\frac{4}{r^2}F_1\,-\frac{6}{r}F'_1\,-4 F''_1\right)
\nonumber\\
&+&  \frac{4}{3 r^2}F_1^3\, .
\end{eqnarray}

We observe that the equation for $F_3^{Skyrme}(t,r)$ in gravitating Skyrme model is a generalization of formulae obtained for wave maps and flat Skyrme model (see $h$--term in equation (18) in paper \cite{BCR2}). Both expressions (\ref{Box_f3}) and (\ref{Box_f3wave}) are linear inhomogeneous wave equations and the only difference is in the form of the source term. Comparing Skyrme and wave maps case we observe that the difference appears in the additional inhomogenity in Skyrme case which we denoted as \textbf{b} - term. We also immediately  see that if we drop \textbf{b} - term in the equation (\ref{Box_f3}) we will get the same expression for the tail as it was obtained for the $\ell=1$ wave maps model \cite{BCRZ}. This expression reads:
\begin{equation}
\label{F3tail} F_3(t,r) = \frac{r}{(t^2-r^2)^2} \left[\alpha C_1 + \mathcal{O}\left( \frac{1}
{t}\right)\right]\,,
\end{equation}
where
\begin{equation}
\label{C1} C_1 = \frac{8}{3} \int \limits_{-\infty}^{+\infty} \left(a''(s)\right)^2 a(s)\, ds\,.
\end{equation}

This form of the expression for $F_3(t,r)$ leads to the following formula for late-time tail at future timelike infinity (i.e. $r=const, t\rightarrow \infty$) \cite{BCRZ}:
\begin{equation}
\label{timelike} F_3(t,r) \simeq \alpha C_1 r t^{-4}.
\end{equation}
We are intrested in calculating this correction caused by  \textbf{b} - term. As it was remarked by Bizo\'n et al. \cite{BCR2}, this term is of lower order in comparison with other terms which contribute to the formula (\ref{F3tail}). Therefore we may expect, that the contribution from \textbf{b} - term will generate higher order corrections which, in principle, are of the form $A/t^{\gamma}$, with $\gamma>4$. To check this and eventually estimate the power--law index  $\gamma$ of possible sub-leading contribution we have solved numerically the equation $\Box F_3^b=\textbf{b}$ and we assume the analytical expression for $F_1$  given by this equation (\ref{f1}). The results of these calculations are plotted in Fig. 1.

\begin{figure}[ht]
\begin{center}
\includegraphics[width=0.6\textwidth]{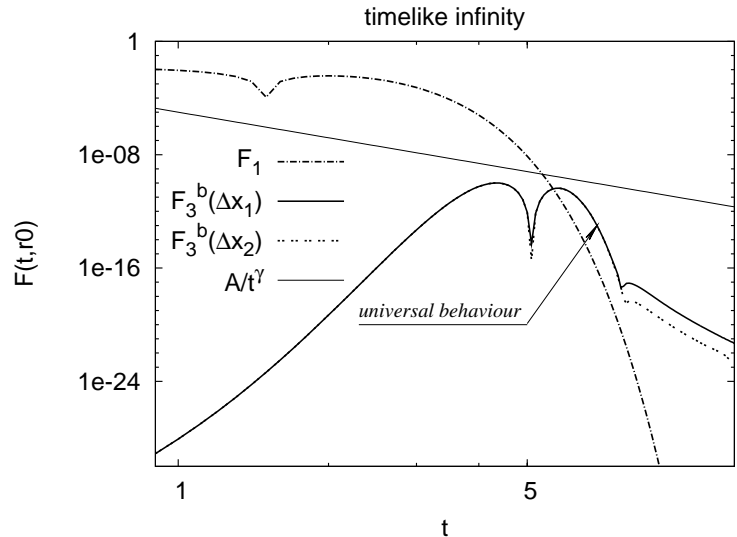}
\end{center}
\caption{\small{The log-log plot of different contributions to the $F(t,r_0)$ for fixed $r_0=5$ as a function of time. We see that $F_3^b$ solution decays faster than any power at future timelike infinity; for a reference we have also plotted a power--law with power--law index $\gamma = 6$.
}} \label{fig.artifact}
\end{figure}


In this figure we have plotted two components of full $F$ solution -- the $ F_1 $ part resulting from the generating function of the form: $a(x)= \varepsilon\, exp(-x^2)$ and the third order correction $F_3^b$ generated by \textbf{b} - term . In fact for the $F_3^b $ we have plotted two curves corresponding to two different resolutions used in numerical calculations. For both contributions we observe  a rising part, which depends on the initial data and falling part, which is more universal. As the generating function $a(x)$ is effectively  the function with compact support, we see that for long times the $F_1$ part of the signal vanishes with time
$ \sim exp(-t^2)$. The main contribution to the tail comes from third-order perturbation of $F$ (see formula (\ref{Box_f3})); to make the figure more transparent we do not plot it here. The most interesting curve in this plot shows the $F_3^b$ component. It is rising part is also initial data dependent. On its falling part ($t>6$) its goes like $F_1$ component --  decreases faster than any power.

For some larger times ($t\geq 7)$, the shape of the curve $F_3^b$  seems to be a power--law. However in our opinion this is not a real effect, but rather an artifact of our numerical procedure coming from a sort of "ghost potential"  (see \cite{Ching} for details). There are two reasons which support such hypothesis. First - this part of the $F_3^b$ curve is very steep (for comparison see a power--law with the power-law index $ \gamma =6$). It is doubtful if this curve may represent a tail.
In addition, the shape of the falling part of $F_3^b$ curve depends on the numerical resolution -- the better resolution the larger part of this curve exhibits the universal behaviour (i.e. decreases faster than any power). This is a typical situation in the case of the "ghost potential".

Summing this up - the numerical calculations suggest, that the contribution of the \textbf{b} - term falls faster than any power - so probably it is equal zero.

To verify this hypothesis we estimate the correction resulting from the \textbf{b} -- term analytically. To do that we solve the equation $\Box F_3^b = b$  by applying the standard Duhamel formula for solving an inhomogeneous wave equation  $\Box F=N(t,r)$ with zero initial data
\begin{equation}
\label{duh} F(t,r)= \frac {1} {2 r} \int \limits_{0}^{t} d\tau \int
\limits_{|t-r-\tau|}^{t+r-\tau} \rho P_{\ell}(\mu)  N(\tau,\rho) d\rho\,.
\end{equation}
Here $P_{\ell}(\mu)$ are Legendre polynomials, in our model $\ell=1$. Using null coordinates
$\eta=\tau-\rho$ and $\xi=\tau+\rho$ and denoting the \textbf{b} kernel by $K(F)$ ,


\begin{eqnarray}
\label{abbK}
K(F) &=& \frac{2}{r^4}\left(F^3 - 2 r F^2\,F' + r^2\,F\,(F'^2-\dot{F}^2)\right),
\end{eqnarray}
we obtain:
\begin{equation}
\label{f3} F_3^b(t,r) = \frac {1} {8 r} \int\limits_{|t-r|}^{t+r} d\xi \int \limits_{-\xi}^{t-r}
(\xi-\eta) P_{\ell}(\mu) K(F_1(\xi,\eta)) d\eta\,,
\end{equation}
where $ \mu=(r^2+(\xi-t)(t-\eta))/r(\xi-\eta)$.
We will assume that the initial data $F_1(t,r)$ are compactly supported, i.e. they vanish outside a ball of some radius $R$. Therefore for $t>r+R$ we can drop  the advanced part of $F_1(t,r)$.  We also change the order of integration in (\ref{f3}) thus we get:
\begin{equation}
\label{f3(2)} F_3^b(t,r) = \frac {1} {8 r} \int \limits_{-\infty}^{\infty} d\eta \int
\limits_{t-r}^{t+r} (\xi-\eta) P_{\ell}(\mu)
K(F^{ret}_1(\xi,\eta)) \, d\xi \,.
\end{equation}

To calculate (\ref{f3(2)}) and obtain the estimate of a $F_3^b$ in timelike infinity, we use the following identity (see paper \cite{BCR5}):
\begin{eqnarray}\label{maF3}
\int \limits_{t-r}^{t+r} d\xi \, \frac {P_{\ell} (\mu)} {(\xi-\eta)^{n}} &=&  (-1)^\ell \frac
{2(n-2)^{\underline{\ell}}} {(2\ell+1)!!} \, \frac {r^{\ell+1}(t-\eta)^{n-\ell-2}} {[(t-\eta)^2 -
r^2]^{n-1}} \, F \left( \left. \begin{array}{c} \frac {\ell+2-n} {2}, \, \frac {\ell+3-n} {2}  \\
\ell + 3/2
\end{array} \right| \left( \frac {r} {t - \eta} \right)^{2} \right)\,
\nonumber\\
&=&
(-1)^\ell \frac{2(n-2)^{\underline{\ell}}}{(2\ell+1)!!} \, \frac {r^{\ell+1}}{t^{\ell+n}} \left( 1 + (\ell+n) \frac{\eta}{t} + \mathcal{O} \left(\frac{1}{t^2} \right) \right)\,.
\end{eqnarray}

From (\ref{f1}) we have
\begin{eqnarray}
\label{f11} F_1 (t,r) =\frac{1}{r} \left(a'(u) +\frac{a(u)}{r} \right)\,,
\end{eqnarray}
\begin{eqnarray}
\label{f1dot} \dot{F}_1(t,r) =\frac{1}{r} \left(a''(u) +\frac{a'(u)}{r}\right)\,,
\end{eqnarray}
 \begin{eqnarray}
\label{f1prim} F_1' (t,r) =-\frac{1}{r} \left(a''(u) +\frac{2 a'(u)}{r} + \frac{2 a(u)}{r^2} \right)\,.
\end{eqnarray}

Substituting (\ref{f11})-(\ref{f1prim}) into (\ref{f3(2)}) and expanding the function $K$ in inverse powers of $\rho=(\xi-\eta)/2$ we get:
\begin{eqnarray}
\label{F3(3)} F_3^b(t,r) &=& \frac {2^5} {r} \int \limits_{-\infty}^{+\infty} d\eta \int
\limits_{t-r}^{t+r} d\xi \, \frac {P_1(\mu)} {(\xi-\eta)^5}  \left[ \frac{2}{3}\frac {d} {d \eta} \left( a'^3(\eta)\right)
\right.
\nonumber\\
&+&
\left.
\frac{1}{\xi-\eta}\left(3 a'^3(\eta)+5 \frac{d}{d \eta} (a'^2(\eta)a(\eta))\right)  + \mathcal{O} \left( \frac{1} {(\xi-\eta)^2} \right)   \right]\,.
\end{eqnarray}
Performing the inner integral over $\xi$ in (\ref{F3(3)}) and using the identity (\ref{maF3}) we obtain

\begin{eqnarray}
\label{F3(31)} F_3^b(t,r) &=&-  2^7 r \int \limits_{-\infty}^{+\infty}d\eta \left[ \frac{1}{t^6}\left(\frac{1}{3}\frac {d} {d \eta} \left( a'^3(\eta)\right)\right) + \frac{1}{t^7}\left(2\frac{d}{d \eta} (\eta a'^3(\eta))+\frac{10}{3}\frac{d}{d \eta}(a'^2(\eta)a(\eta))\right)  + \mathcal{O} \left( \frac{1} {t^8} \right)   \right]\, .
\end{eqnarray}


We have obtained an expression which gives the expansion of the result in inverse powers of $t$. The factors multiplying this inverse powers of $t$ are integrals of total derivatives of expressions which vanish at the integral boundaries, therefore they are equal zero. As a result we get the following estimation for $F_3^b$  at future timelike infinity:
\begin{equation}
F_3^b= \mathcal{O}\left(\frac{1}{t^8}\right)\,,
\end{equation}

i.e. we have demonstrated that $F_3^b$ does not contains terms of the form $A/t^\gamma$ with $\gamma <7$. 
We would like to stress here that in our analysis we considered only two terms to show their cancellation. However, this method may be extended to higher orders. To make it possible we have to expand the identity (\ref{maF3}) to higher orders in $1/t$. If we do that and proceed in the same way we will get the cancelations of higher order coefficients, what is in agreement with the hypothesis that $F_3^b$ asymptotically tends to $zero$.

In summary -- analytical results support the hypothesis stated on the basis of numerical results. Both analytical and numerical results are compatible with the fact that the \textbf{b} -- term does not contribute to the asymptotic expression for the tail. As a results the Skyrme model and $\ell = 1$ wave map model are the examples of models which although are different, have the same long time asymptotics.

\section{Numerics}
\label{sec:numerics}

To verify analytical prediction for the tails obtained in the previous section we have performed numerical studies of long-time asymptotics in the Einstein-Skyrme model.
To do that we have solved numerically the equations (\ref{momentum})-(\ref{wave}) with initial data described below.
For solving evolutional equations we have used method of lines with 5-point, fourth order accurate spatial discretization. We have solved the resulting ODE's with fourth order Runge-Kutta method.
To solve the costraints, i.e. hamiltonian constrain (\ref{hamilton}) and slicing condition (\ref{delta}) we have also used fourth order Runge-Kutta method.
Here we need the  values of some functions out of the grid  -- we have obtained them using spline interpolation.
To ensure regularity at the origin we impose the boundary conditions $F(t,r=0) \sim r$ and $P(t,r=0)\sim r$.
To avoid the contamination of results by parts of the solution reflected from outer boundary we have used the size of the grid big enough, so the solution stops before the reflected signal reaches the observation point. Finally, to suppress the accumulation of round-off errors in late times we have used quadrupole precision.
In our calculations we have used the initial data generated by the function (see (\ref{id})-(\ref{f1})) so we get: $a(x)=\varepsilon \, exp(-x^2)$ for different values of $\varepsilon$. We have started with comparing the behaviour of solutions in the Einstein-Skyrme model and $\ell=1$ wave maps model. We have prepared the same initial data and evolved them in both models. The results of these simulations are shown in Fig. 2. In the left panel of this figure we plot solutions $F(t,r_0)$  for a fixed observation point $r_0$ as a function of time whereas in the right panel we plot solutions $F(u,v_0)/r$  for fixed value of $v_0$ as a function of $u$.

\begin{figure}[ht]
\begin{tabular}{cc}
\includegraphics[width=0.42\textwidth]{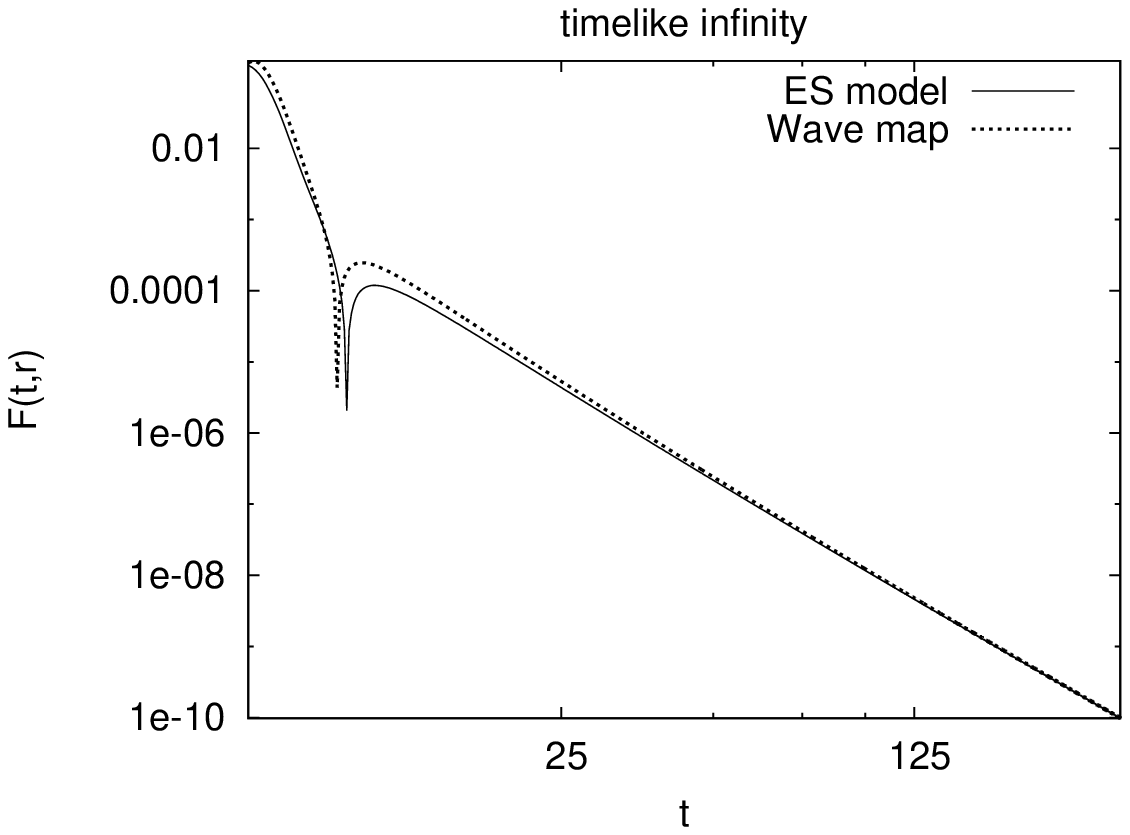}
&
\includegraphics[width=0.42\textwidth]{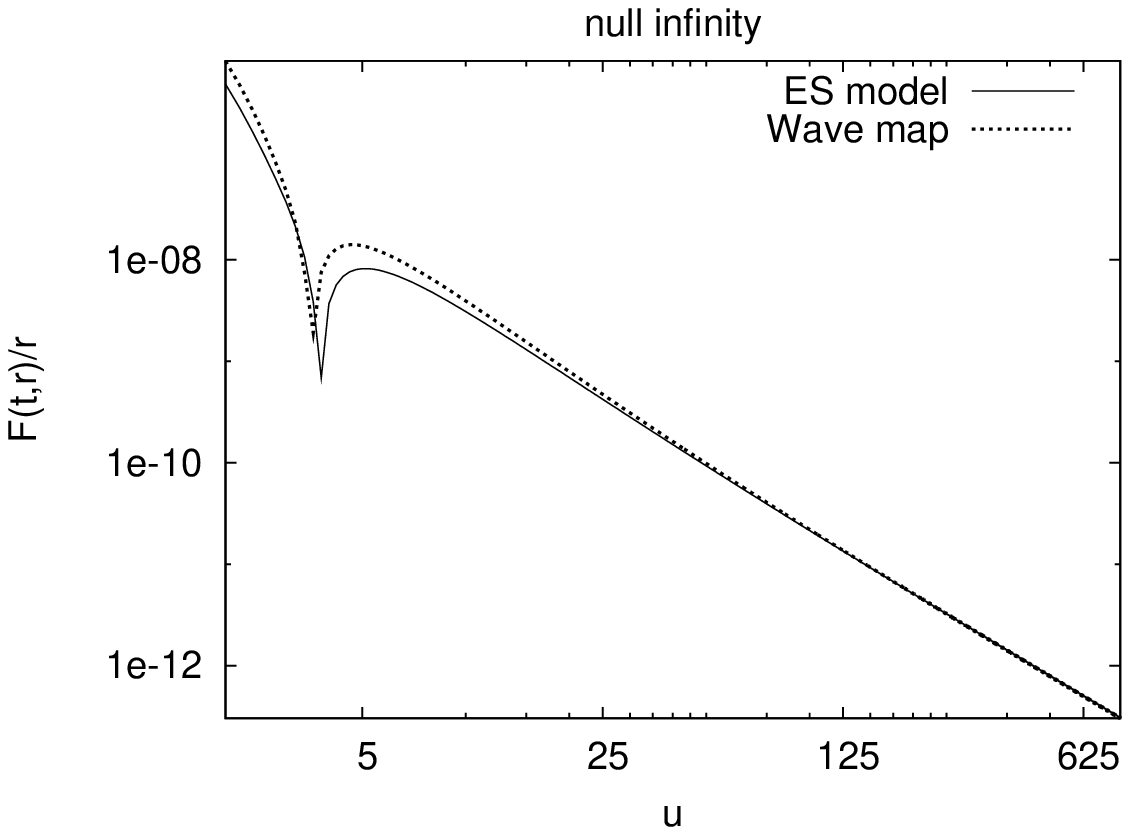}
\\
\end{tabular}
\caption{\small{Left panel: The log-log plot of $F(t,r_0)$ vs. $t$ for fixed $r_0=5$. Right panel: The log-log plot of  $F(u,v_0)/r$ for fixed large advanced time $v_0=t+r=1000$ as the function of retarded time $u=t-r$. In both model we use $\alpha=0.03$ and $\varepsilon=1.0$.}} \label{fig.compare}
\end{figure}

We observe from above figure differencies at the begining of the evolution and in the intermediate asymptotics but for large-times they dissapear. It means that these two models are different but have the same long-time asymptotics.

In Fig.3 we plot $F(t,r)$ in self--gravitating Skyrme model with $\alpha=0.03$ for three different values of $\varepsilon$. We see that on log-log plots the late-time tails are clearly seen as straight lines.
\begin{figure}[ht]
\begin{tabular}{cc}
\includegraphics[width=0.45\textwidth]{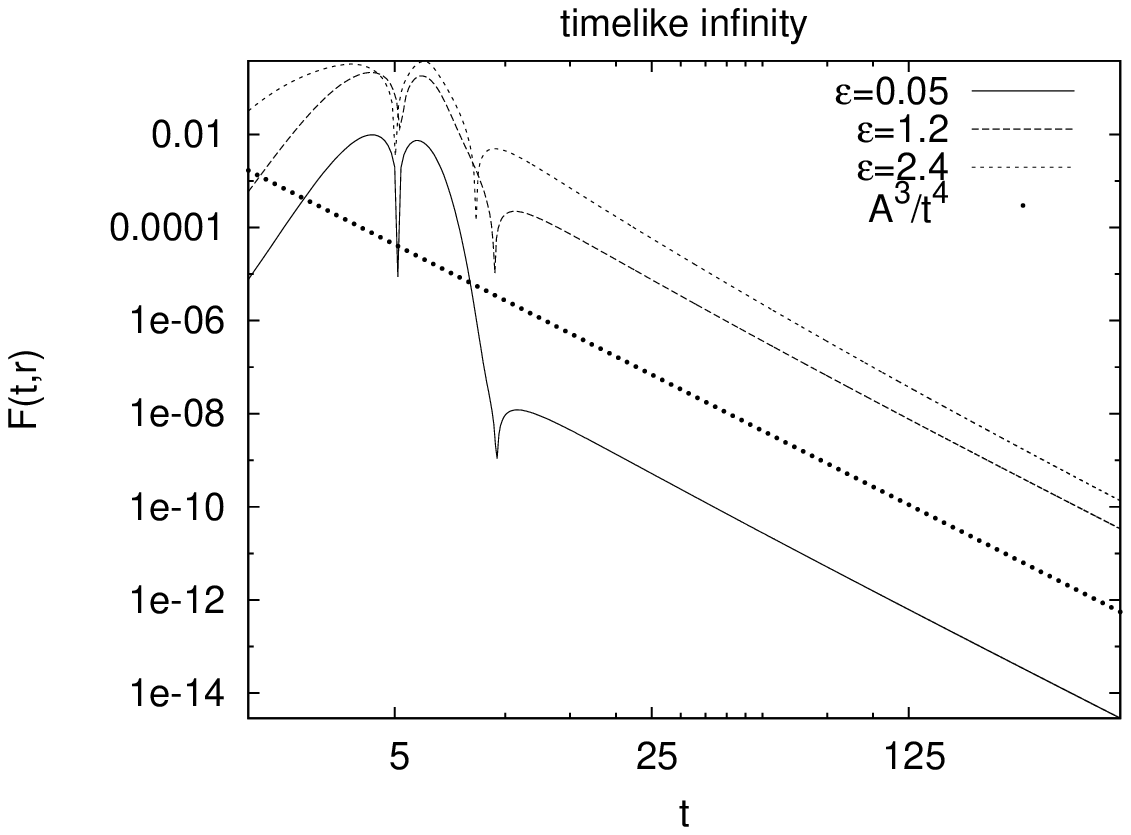}
&
\includegraphics[width=0.45\textwidth]{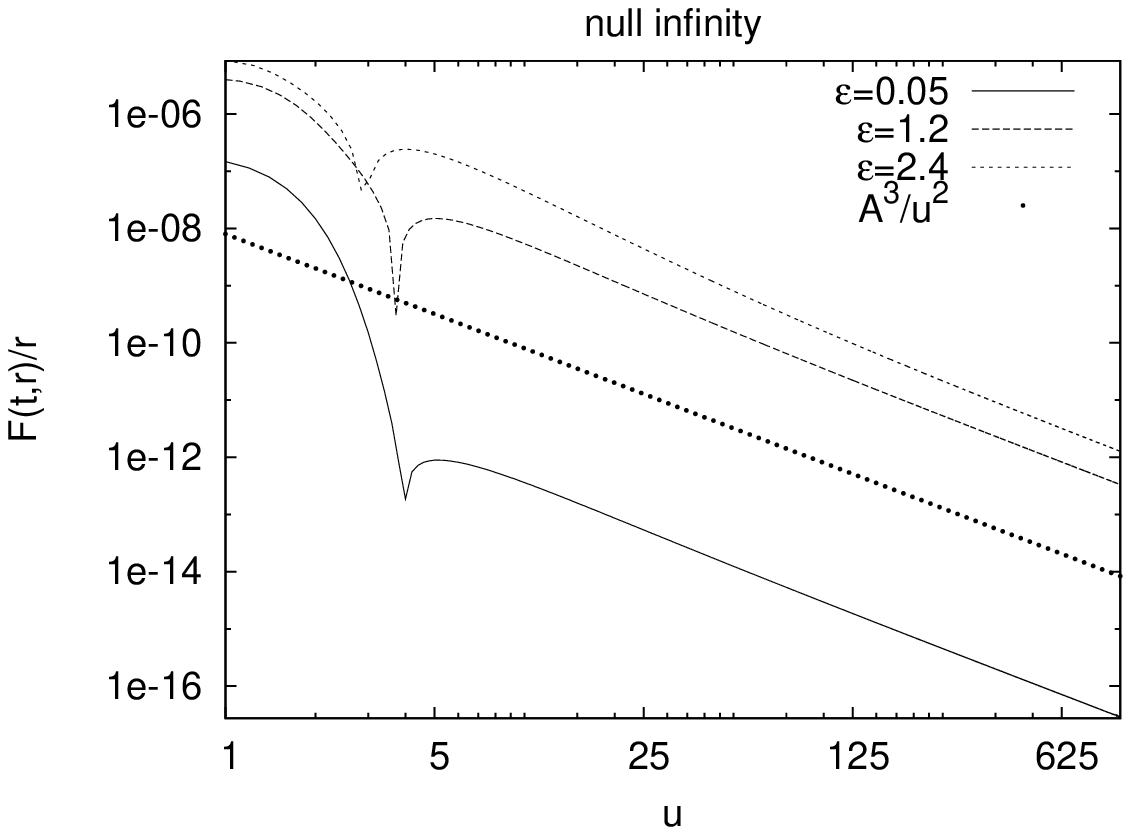}
\\
\end{tabular}
\caption{\small{Left panel: The log-log plot of $F(t,r)$ for fixed r=5. Right panel: The log-log plot of  $F(t,r)/r$ for fixed large advanced time $v=t+r=1000$ as the function of retarded time $u=t-r$. In both panels (dotted line) we see that solutions starting from small initial data decay as $t^{-4}$ at future timelike infinity and as $u^{-2}$ at future null infinity}} \label{fig.null}
\end{figure}

To obtain the parameters of the tails we should use the following formula:
\begin{equation}
\label{tail_fit}
F(t,r) = A t^{-\gamma} \exp \left(B/t + C/t^2 \right).
\end{equation}

For the comparison with numerical data it is convenient to define the local power index (hearafter LPI) defined as follows \cite{Burko&Ori}:
\begin{equation}
n(t,r)=-t\frac{\dot{F}(t,r)}{F(t,r)}.
\end{equation}
For the assumed form parametrising the tail (\ref{tail_fit}) we get the following expression for the LPI:
\begin{equation}
n(t,r)=\gamma+\frac{B}{t}+\frac{2C}{t^2}.
\end{equation}

In Fig. 4 we plot LPI  at $r=5$ as a function of $1/t$. All curves in this figure correspond to small initial data. We see that all lines approach the same power-law index $\gamma=4$ at the future timelike infinity, so numerical data confirm analytical prediction for the decay rate see equation (\ref{timelike}).

\begin{figure}[ht]
\begin{center}
\includegraphics[width=0.45\textwidth]{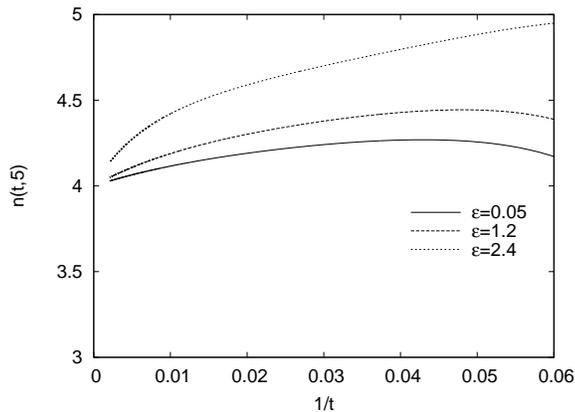}
\end{center}
\caption{\small{The local power index n(t,5) as a function of 1/t.
 }} \label{fig.lpi}
\end{figure}
In Fig.~5 we plot $\varepsilon^{-3} F(t,r)$ as a function of initial amplitude. According to the analytical prediction the late--time behaviour of this quantity does not depend on the magnitude of initial data. We may observe that for not-too-large initial data this is really the case.

\begin{figure}[ht]
\begin{tabular}{cc}
\includegraphics[width=0.45\textwidth]{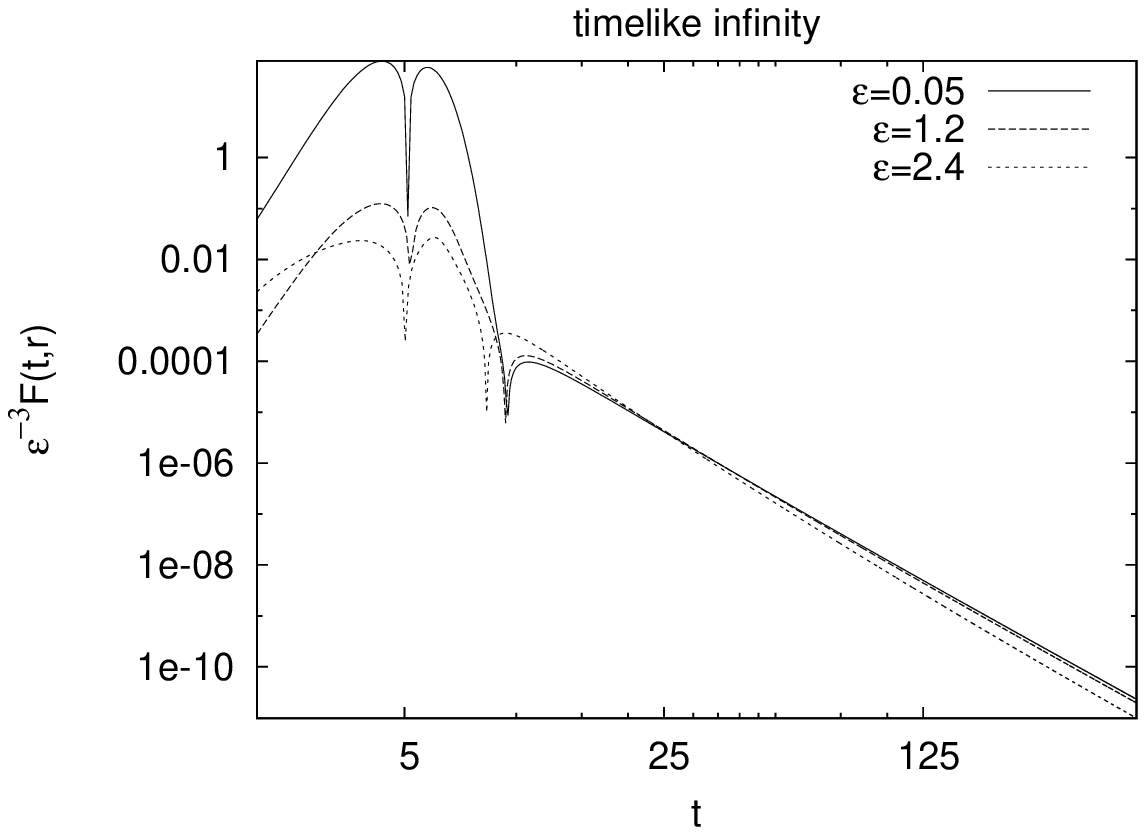}
&
\includegraphics[width=0.45\textwidth]{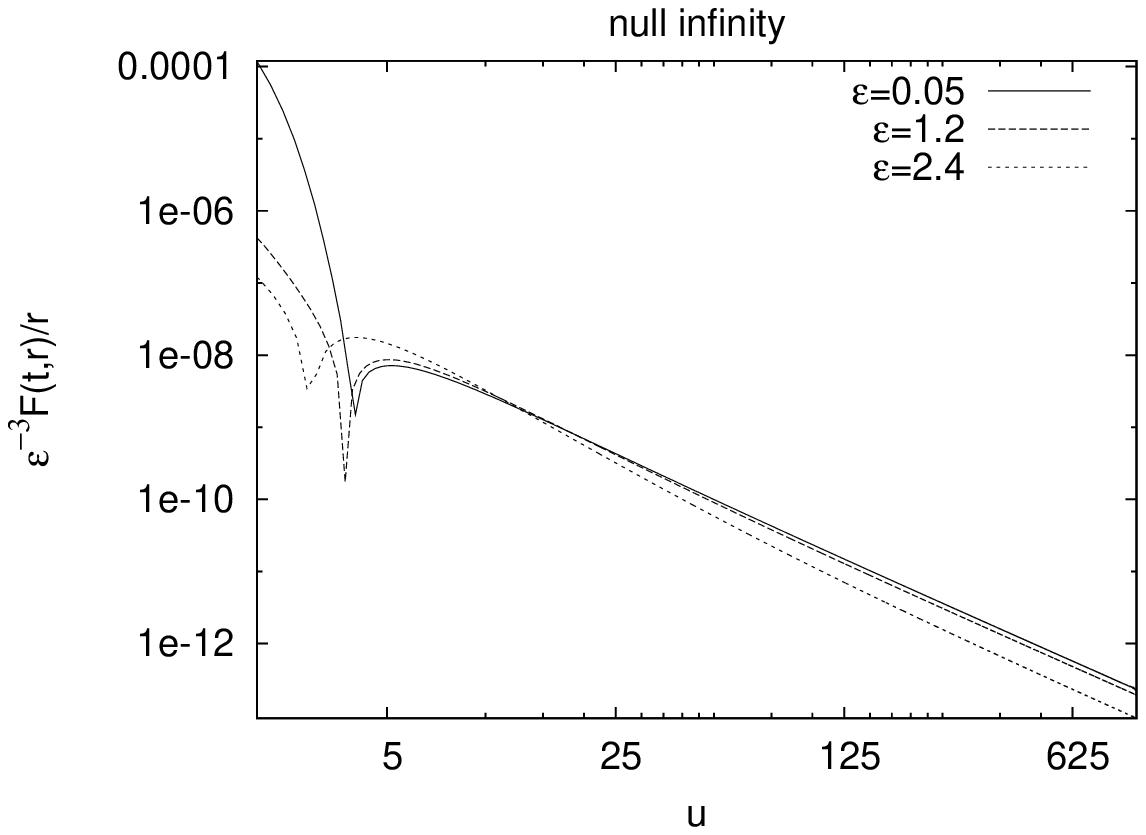}
\\
\end{tabular}
\caption{\small{Left panel: The log-log plot of $\varepsilon^{-3} F(t,r)$ vs time for fixed r=5. Right panel: The log-log plot of  $\varepsilon^{-3} F(t,r)/r$ for fixed large advanced time $v=t+r=1000$ as the function of retarded time $u=t-r$.}} \label{fig.null_ep}
\end{figure}

\textbf{Summary: }
Using a $nonlinear$ perturbation method, we have demonstrated that leading order formulae for late-time tails in $ES$ model are exactly the same as those obtained for wave maps (if we drop this correction \textbf{b} -- term). In other words, these two models albeit, in principle, different have the same long-time asymptotics and it is seen in Fig. 2. We have also checked hypothesis that this correction \textbf{b} -- term disapeared faster than any other power so that  $F_3^b$ (see equation (\ref{b})) is probably equal zero (see Fig. 1). We verified numerically the power-law index for the tail $t^{-4}$ at future timelike infinity and $u^{-2}$ at future null infinity. We also confirmed that for the Einstein-Skyrme model we will get the same analytical results in the leading order for the tails as was obtained by Bizo\'n et al. \cite{BCRZ}, because this \textbf{b} - term  does not contribute to the asymptotic expression for the tail.

\vskip 0.2cm \noindent \textbf{Acknowledgments:} I am greatly indebted to Andrzej Rostworowski and Tadeusz Chmaj for discussions and remarks. We acknowledge support by the MNII grants NN202 079235.


\begin{thebibliography}{}

\bibitem{TCh} Tadeusz Chmaj, private communication.

\bibitem{SZ} Stanis{\l}aw Zaj\c{a}c, Acta Phys. Polon. \textbf{B} 40, 1617-1628 (2009).

\bibitem{Ching} Ching et al., Phys. Rev. D \textbf{52}, 2118  (1995).

\bibitem{BCR1} P. Bizo\'n, T. Chmaj, and A. Rostworowski, math-ph/0701037

\bibitem{BCRZ} P. Bizo\'n, T. Chmaj, A. Rostworowski and S. Zaj\c{a}c, Class. Quantum Grav. 26, 225015 (2009).

\bibitem{BCR2} P. Bizo\'n, T. Chmaj, and A. Rostworowski, Phys. Rev. D \textbf{75}, 121702(R) (2007).

\bibitem{BCR3} P. Bizo\'n, T. Chmaj, and A. Rostworowski, Class. Quantum Grav. 24, F55 (2007).

\bibitem{BCR4} P. Bizo\'n, T. Chmaj, and A. Rostworowski, Phys. Rev. D \textbf{76}, 124035 (2007).

\bibitem{BCR5} P. Bizo\'n, T. Chmaj, and A. Rostworowski, Phys. Rev. D \textbf{78}, 024044 (2008).

\bibitem{Skyrme} T.H.R. Skyrme, Proc. R. Soc. \textbf{A 260}, 127 (1961).

\bibitem{Burko&Ori} L.M. Burko and A. Ori, Phys. Rev. D \textbf{56}, 7820 (1997).


\end{thebibliography}
\end{document}